\documentclass{article}

\begin{document}

\title{A stabilized Kuramoto-Sivashinsky system}
\author{Boris A. Malomed \\
Department of Interdisciplinary Studies, Faculty of Engineering,\\
Tel Aviv University, Tel Aviv 69978, Israel\\
Bao-Feng Feng \\
Department of Mathematics, the University of Kansas,\\
Lawrence, KS 66045, U.S.A. \and Takuji Kawahara \\
Department of Aeronautics and Astronautics, \\
Graduate School of Engineering,\\
Kyoto University, Sakyo-ku, Kyoto 606-8501, Japan}
\maketitle

\begin{center}
\textbf{Abstract}
\end{center}

A model consisting of a mixed Kuramoto - Sivashinsky - Korteweg - de Vries
equation, linearly coupled to an extra linear dissipative equation, is
proposed. The model applies to the description of surface waves on
multilayered liquid films. The extra equation makes its possible to
stabilize the zero solution in the model, thus opening way to the existence
of stable solitary pulses (SPs). By means of the perturbation theory,
treating dissipation and instability-generating gain in the model (but not
the linear coupling between the two waves) as small perturbations, and
making use of the balance equation for the net momentum, we demonstrate that
the perturbations may select two steady-state solitons from their continuous
family existing in the absence of the dissipation and gain. In this case,
the selected pulse with the larger value of the amplitude is expected to be
stable, provided that the zero solution is stable. The prediction is
completely confirmed by direct simulations. If the integration domain is not
very large, some pulses are stable even when the zero background is
unstable. An explanation to the latter finding is proposed. Furthermore,
stable bound states of two and three pulses are found numerically.


\newpage

\section{Introduction}

The Kuramoto~-~Sivashinsky (KS) equation is a well-known model of
one-dimensional turbulence, which was derived in various physical contexts,
including chemical-reaction waves \cite{Kuramoto}, propagation of combustion
fronts in gases \cite{Sivashinsky}, surface waves in a film of a viscous
liquid flowing along an inclined plane \cite{Nepom}, patterns in thermal
convection \cite{book}, rapid solidification \cite{Philip}, and others. It
has the form 
\begin{equation}
u_{t}+uu_{x}=-\alpha u_{xx}-\gamma u_{xxxx},  \label{KS}
\end{equation}
where $\alpha >0$ and $\gamma >0$ are coefficients accounting for the
long-wave instability (gain) and short-wave dissipation, respectively.

A generalized form of the KS equation contains a linear dispersive term
borrowed from the Korteweg~-~de~Vries (KdV) equation, 
\begin{equation}
u_{t}+uu_{x}+u_{xxx}=-\alpha u_{xx}-\gamma u_{xxxx}\,.  \label{Benney}
\end{equation}
As well as the KS equation proper, the generalized equation (\ref{Benney})
applies to the description of surface waves on flowing liquid films \cite
{Ivansky}, and it also serves as a general model which allows one to study
various nonlinear dissipative waves \cite{Takuji}-\cite{KSpulse2}. In
particular, a subject of considerable interest was the study of
solitary-pulse (SP) solutions to both the KS equation \cite{KSpulse} and Eq.
(\ref{Benney}) \cite{Takuji2,KSpulse2}. By means of numerical methods, it is
possible to find a vast family of SP solutions to Eqs. (\ref{KS}) and (\ref
{Benney}) in the form $u(x,t)=u(x-st)$ with a constant velocity $s$.
However, it is obvious that all these solutions are unstable in an
infinitely long system, as the zero solution, into which SP goes over at $%
|x-st|\rightarrow \infty $, is unstable in both equations.

It is an issue of a principal interest to find a physically relevant model
which combines the dissipative and dispersive features, and simultaneously
supports \emph{stable} SPs. It appears that the simplest possibility to
construct such a model is to couple Eq. (\ref{Benney}) to an extra linear
stabilizing equation, arriving at a system 
\begin{eqnarray}
u_{t}+uu_{x}+u_{xxx} &=&-\alpha u_{xx}-\gamma u_{xxxx}+\epsilon _{1}v_{x},
\label{1} \\
v_{t}+cv_{x} &=&\Gamma v_{xx}+\epsilon _{2}u_{x},  \label{2}
\end{eqnarray}
where the dissipative parameter (effective diffusion coefficient) $\Gamma >0$
accounts for the stabilization (see below), and $c$ is a group-velocity
mismatch between the two wave modes. The coupling parameters $\epsilon _{1}$
and $\epsilon _{2}$ must have the same sign (otherwise the coupling gives
rise to an instability), while their magnitudes may be different. However,
it is always possible to make them equal, $\epsilon _{1}=\epsilon _{2}\equiv
\epsilon $, by means of an obvious rescaling of $u$ and $v$. Then, using the
remaining scaling invariance of the equations, it is possible to set $%
\epsilon \equiv 1.$ Thus, we will be dealing with a system containing four
irreducible parameters, 
\begin{eqnarray}
u_{t}+uu_{x}+u_{xxx}-v_{x} &=&-\alpha u_{xx}-\gamma u_{xxxx},  \label{u} \\
v_{t}+cv_{x}-u_{x} &=&\Gamma v_{xx}.  \label{v}
\end{eqnarray}
Note that the system conserves two ``masses'', 
\begin{equation}
M=\int_{-\infty }^{+\infty }u(x)dx,\,\,N=\int_{-\infty }^{+\infty }v(x)dx.
\label{mass}
\end{equation}

The system of equations (\ref{1}) and (\ref{2}) can find their natural
physical realization as a model describing coupled surface and interface
waves in a two-layered flowing liquid film, cf. the similar interpretation
of the single equation (\ref{KS}) or Eq. (\ref{Benney}) mentioned above. In
particular, the linear coupling via the first derivatives is the same as in
known models of coupled internal waves propagating in multi-layered fluids 
\cite{Grimshaw}. Then, the linear dissipative equation (\ref{2}) implies
that the substrate layer is essentially more viscous than the upper one. In
fact, the additional equation (\ref{v}) may also be nonlinear, but it can be
checked that inclusion of the nonlinear term $vv_{x}$ into this equation
does not produce any conspicuous difference, therefore we focus on the
simplest model (\ref{u}), (\ref{v}), which provides for the stabilization of
SPs.

The system of equations (\ref{u}) and (\ref{v}) is qualitatively similar to
a system of linearly coupled Ginzburg-Landau (GL) equations describing
propagation of localized pulses in an fiber-optic core equipped with
distributed gain, which is linearly coupled to an extra lossy core that
provides for the stability of the pulses \cite{we,Atai2,further} (such
double-core systems have recently become available to experimental studies,
and they have very promising features for applications to optical
communications, see a short overview in Ref. \cite{optics}). The most
fundamental version of this GL system is that in which the extra stabilizing
equation is also linear, cf. Eq. (\ref{v}); in that case, SP solutions can
be found in an exact analytical form, and they are stable in a certain
parametric region \cite{Atai2}.

In this work, we will find \emph{stable} SPs in the system (\ref{u}), (\ref
{v}), which appears to be the first example of stable pulses in a model of
the KS type. In section 2, we analyze the stability of the zero solution,
which, as it was mentioned above, is a necessary condition for the stability
of SPs in the infinitely long system. In section 3, an analytical
perturbation theory for the pulses is developed, which is based on treating
the gain and dissipation constants $\alpha $, $\gamma $, and $\Gamma $ in
Eqs. (\ref{u}) and (\ref{v}) as small parameters (while the group-velocity
mismatch $c$ need not be small). In the zero-order approximation, $\alpha
=\gamma =\Gamma =0$, Eqs. (\ref{u}) and (\ref{v}) have a one-parameter
family of exact soliton solutions. Using the known approach based on the
balance equation for the momentum \cite{Takuji}, we demonstrate that the
combination of the perturbation terms in Eqs. (\ref{u}) and (\ref{v}) may
select one or two stationary pulses out of the continuous family existing in
the zero-order approximation. As is known \cite{we}, the existence of two
different SP solutions is a necessary condition for the stability of one of
them, the second pulse (the one with smaller amplitude) being unstable, as
it plays the role of a separatrix between attraction domain of the zero
solution and stable pulse.

In section 4, we present results of direct numerical simulations of the full
system (\ref{u}), (\ref{v}), which demonstrate that stable SPs exist indeed.
In fact, simulations sometimes produce stable pulses even in the case when
the zero solution is not stable. This stability extension may be explained
by the finite size of the simulation domain. Moreover, stable bound states
(BSs) of two and three pulses are found numerically too.

\section{The stability of the zero solution}

As it was explained above, it is necessary to investigate the stability of
the trivial solution, $u=v=0$, before the consideration of pulses. To this
end, we substitute into the linearized equations (\ref{u}) and (\ref{v}) a
perturbation in the form $u\sim \exp \left( ikx+\lambda t\right) $, $\,v\sim
\exp \left( ikx+\lambda t\right) $, where $k$ is an arbitrary real wave
number of the perturbation, and $\lambda $ is the corresponding instability
growth rate, which leads to a dispersion equation, 
\begin{equation}
\left( \lambda -ik^{3}-\alpha k^{2}+\gamma k^{4}\right) \left( \lambda
+ick+\Gamma k^{2}\right) +k^{2}=0.  \label{lambda}
\end{equation}
The stability condition states that both solutions of the quadratic equation
(\ref{lambda}) must satisfy the inequality $\mathrm{Re\,}\left[ \lambda (k)%
\right] \leq 0$ at all the real values of $k$.

For $k\rightarrow 0$, solutions to Eq. (\ref{lambda}) can be found in the
form of an expansion 
\begin{equation}
\lambda (k)=i\lambda _{1}k+\lambda _{2}k^{2}+...\,,  \label{expansion}
\end{equation}
where $\lambda _{1}=\left( -c\pm \sqrt{c^{2}+4}\right) /2$, and 
\begin{equation}
\lambda _{2}=\frac{-(\Gamma -\alpha )\sqrt{c^{2}+4}\pm (\Gamma +\alpha )c}{2%
\sqrt{c^{2}+4}}\,,  \label{lambda2}
\end{equation}
the sign $\pm $ being the same in $\lambda _{1}$ and $\lambda _{2}$. As $%
\lambda _{1}$ is always real, at the first order the expansion (\ref
{expansion}) implies neutral stability. The expression (\ref{lambda2})
yields a necessary condition for the stability of the zero solution at the
second order of the expansion, $\mathrm{Re\,}\lambda _{2}\leq 0$, which can
be cast into a form 
\begin{equation}
\Gamma -\alpha \geq \sqrt{\alpha \Gamma }\left| c\right| .  \label{necessary}
\end{equation}
In the particular case $c=0$, the condition (\ref{necessary}) amounts to $%
\Gamma >\alpha $, which has a simple meaning: the stabilizing diffusion
coefficient in Eq. (\ref{v}) must be larger than the instability-driving
``anti-diffusion'' (gain) coefficient in Eq. (\ref{u}). A very similar
necessary stability condition is known in the above-mentioned system of
coupled GL equation describing a dual-core optical fiber with one active and
one passive cores \cite{we,Atai2}.

Comprehensive analysis of the zero-solution stability was performed by means
of numerical solution of the dispersion equation (\ref{lambda}). It was
found that the full stability condition does not amount to the inequality (%
\ref{necessary}) (i.e., it may happen that $\mathrm{Re\,}\left[ \lambda (k)%
\right] $ is negative at small $k$, but it takes positive values in some
interval of finite values of $k$). Numerically found stability borders in
the plane of the parameters ($\alpha ,\Gamma $) for a fixed value $\gamma
=0.05$ of the short-wave stabilization parameter in Eq. (\ref{u}) and two
different values of the group-velocity mismatch, $c=0$ and $c=-1$, are shown
by dashed curves in stability diagrams for the pulses displayed in Figs. 1
and 2. In both cases, the zero solution is stable to the left of the
stability border. Note that, for small $\alpha $ and $\Gamma $, the
zero-solution stability region is indeed determined by Eq. (\ref{necessary}%
), but at larger values of $\alpha $ and $\Gamma $ there appear additional
stability restrictions.

\section{The perturbation theory for solitary pulses}

At the zeroth order, setting $\gamma =\Gamma =\alpha =0$ in Eqs. (\ref{u})
and (\ref{v}), but keeping an arbitrary value of $c$, we arrive at a
conservative system consisting of the KdV equation coupled to an extra
linear one, 
\begin{equation}
u_{t}+uu_{x}+u_{xxx}=v_{x},v_{t}+cv_{x}=u_{x}.  \label{zero_order}
\end{equation}
Equations (\ref{zero_order}) have a family of exact two-component soliton
solutions, 
\begin{equation}
u(x,t)=12\eta ^{2}\mathrm{sech}^{2}\left( \eta \left( x-st\right) \right)
,\,v(x,t)=\left( c-s\right) ^{-1}\cdot u(x,t),  \label{soliton}
\end{equation}
where $\eta $ is an arbitrary parameter which determines the soliton's
amplitude and width, and the velocity $s$ takes two different values for
given $\eta $, 
\begin{equation}
s=\frac{1}{2}\left[ \left( c+4\eta ^{2}\right) \mp \sqrt{\left( c-4\eta
^{2}\right) ^{2}+4}\right] .  \label{s}
\end{equation}
It will be more convenient to use, as a parameter of the soliton family, not
the amplitude $\eta $, but rather the relative velocity, 
\begin{equation}
\delta \equiv c-s,  \label{delta}
\end{equation}
in terms of which the amplitude is given by an expression obtained from Eq. (%
\ref{s}), 
\begin{equation}
4\eta ^{2}=c-\delta +1/\delta \,.  \label{eta}
\end{equation}
The range of meaningful values of $\delta $ is restricted by the condition $%
\eta ^{2}>0$.

We have checked by direct simulations of Eqs. (\ref{zero_order}) that all
the soliton solutions (\ref{soliton}) are stable within the framework of the
unperturbed equations (\ref{zero_order}). On the other hand, simulations
also clearly demonstrate that collisions between solitons having different
velocities are inelastic (although not strongly inelastic, see a typical
example in Fig. 3), hence the conservative system (\ref{zero_order}), unlike
the KdV equation proper, is \emph{not} an exactly integrable one. The
nonintegrability of the system (\ref{zero_order}) has also been confirmed by
analysis of its symmetries performed by G. Burde (unpublished).

The next step is to restore the small dissipative and gain perturbations,
getting back from Eqs. (\ref{zero_order}) to Eqs. (\ref{u}) and (\ref{v}).
To this end, we notice that the unperturbed equations (\ref{zero_order})
conserve not only the masses (\ref{mass}) but also the net momentum, 
\begin{equation}
P=\frac{1}{2}\int_{-\infty }^{+\infty }\left( u^{2}+v^{2}\right) dx.
\label{P}
\end{equation}
Following the work \cite{Takuji}, in the first approximation of the
perturbation theory the evolution of the soliton may be described by means
of the \textit{balance equation} for the momentum. Indeed, a consequence of
Eqs. (\ref{u}) and (\ref{v}) is the following exact evolution equation for
the net momentum in the presence of the perturbations: 
\begin{equation}
\frac{dP}{dt}=\int_{-\infty }^{+\infty }\left( \alpha u_{x}^{2}-\gamma
u_{xx}^{2}-\Gamma v_{x}^{2}\right) dx.  \label{dPdt}
\end{equation}

Steady-state SPs are selected by the condition $dP/dt=0$. The right-hand
side of Eq. (\ref{dPdt}) can be explicitly calculated in the approximation
in which $u$ and $v$ are substituted by the expressions (\ref{soliton}).
After a straightforward algebra, the equation $dP/dt=0$ can be cast into the
form of a cubic equation for the relative velocity $\delta $ of the
unperturbed soliton (\ref{soliton}), 
\begin{equation}
5\delta ^{3}+\left( 7\widetilde{\alpha }-5c\right) \delta ^{2}-5\delta -7%
\widetilde{\Gamma }=0,\,\widetilde{\alpha }\equiv \alpha /\gamma ,\widetilde{%
\Gamma }\equiv \Gamma /\gamma .  \label{cubic}
\end{equation}

Roots of Eq. (\ref{cubic}) select SPs that may exist as steady states within
the framework of the perturbation theory. Note that, besides the obvious
condition that physical roots for $\delta $ must be real (they may be both
positive and negative), they must also satisfy a condition that, after the
substitution into Eq. (\ref{eta}), they must produce $\eta ^{2}>0$.
Generally speaking, there may exist up to three physical roots of Eq. (\ref
{cubic}); however, in a vast parametric area considered, we have never
encountered a case when Eq. (\ref{cubic}) would indeed have three physical
roots, while the existence of two physical solutions is quite possible, see
below. As it was mentioned in the Introduction, one may expect that SP may
be stable if \emph{precisely two} different pulses exist, then the one with
the larger amplitude $\eta ^{2}$ has a chance to be stable, while the pulse
with the smaller amplitude is always unstable \cite{we,Atai2}. Indeed, if
there is a stable SP, we are dealing with a bistable system, as the
parameters are chosen so that the zero solution is also stable, see the
previous section. In a bistable system, there should exist a \textit{%
separatrix}, i.e., a border between attraction domains of two stable
solutions, the separatrix itself being an unstable stationary solution. In
the situation with two different stationary SP solutions predicted by the
perturbation theory, the one with the smaller amplitude and larger width is
a natural candidate to the role of the unstable separatrix solution, while
its counterpart with the larger amplitude and smaller width may be stable
(generally speaking, it may be stable in a part of the parametric region
where this situation takes place \cite{we,Atai2}).

Equation (\ref{cubic}) may be simplified in the case $c=0$, if we
additionally assume that both renormalized parameters $\widetilde{\alpha }$
and $\widetilde{\Gamma }$ are large (in fact, we are interested in the case
when $\widetilde{\Gamma }\sim \widetilde{\alpha }^{3}\gg 1$). Then, the term 
$-5\delta $ may be neglected in Eq. (\ref{cubic}), so that it takes the form 
\begin{equation}
5\delta ^{3}+7\widetilde{\alpha }\delta ^{2}-7\widetilde{\Gamma }=0.
\label{simplified}
\end{equation}
In the case $c=0$, the condition $\eta ^{2}>0$ following from Eq. (\ref{eta}%
) also takes a simple form: a physical root is that which belongs to either
of the two intervals, 
\begin{equation}
\delta <-1;\,\,0<\delta <1.  \label{physical}
\end{equation}
With regard to the assumption that $\widetilde{\Gamma }\ $and $\widetilde{%
\alpha }$ are large, it is easy to see that the simplified equation (\ref
{simplified}) always has a real root in the region $\delta >1$, which is
unphysical according to Eq. (\ref{physical}). Two physical roots $\delta <-1$
exist under the condition 
\begin{equation}
\widetilde{\Gamma }<\frac{1}{3}\left( \frac{14}{15}\right) ^{2}\widetilde{%
\alpha }^{3}.  \label{condition}
\end{equation}
Note that this condition does not contradict the necessary condition (\ref
{necessary}) of the stability of the zero solution, which in the present
case ($c=0$) takes the form $\widetilde{\Gamma }>\widetilde{\alpha }$.
Indeed, the latter inequality is compatible with (\ref{condition}), provided
that $\widetilde{\alpha }>\sqrt{3}\left( 15/14\right) \approx \allowbreak
1.\,\allowbreak 856$, which is correct, as we are here dealing with the case
when $\widetilde{\alpha }$ is large. However, the inequality (\ref{condition}%
) is not necessarily compatible with the full stability condition for the
zero solution, see Fig. 1.

In the general case, it is easy to solve Eq. (\ref{cubic}) numerically.
Then, selecting a parametric region in which there are exactly two physical
solutions (which, as it was explained above, is a necessary condition for
the existence of a stable SP), one may identify a narrower region in which
this condition holds and, simultaneously, the zero solution is stable.
Stable pulses may exist only inside that region where both necessary
stability conditions overlap, and direct simulations show that \emph{all}
the pulses belonging to the region are stable indeed, at least in case
displayed in Fig. 1, see details below.

The so defined regions in the parametric plane ($\alpha ,\Gamma $), in which
stable SPs are expected, are displayed, for $\gamma =0.05$, in Figs. 1 and 2
for $c=0$ and $c=-1$, respectively. The condition of the existence of
exactly two different physical solutions for the pulses holds to the right
of the continuous curve in these figures [note that the part of the curve
corresponding to sufficiently large values of $\Gamma $ is well approximated
by the analytical expression (\ref{condition}) obtained above]. At the
points belonging to the continuous curve, the two physical solutions merge
and disappear via a typical tangent (saddle-node) bifurcation.

The same analysis performed for values of the short-wavelength-dissipation
parameter $\gamma $ different from the value $0.05$, for which the results
are presented in Figs. 1 and 2, shows that variation of $\gamma $ produces
little change in terms of the expected SP stability region (generally, the
size of the region increases with $\gamma $). As for the effect of the
group-velocity mismatch $c$, we have found that the stability region quickly
shrinks with the increase of $c$ when $c$ is positive, and there is no
stability region at $c>c_{\mathrm{cr}}$, where $c_{\mathrm{cr}}$ is
slightly larger than $0.3$. In that case, the areas in which, respectively,
the zero solution is stable, and there are two different stationary SPs, do
not overlap. To illustrate this point, a very narrow stability region in the
($\alpha ,\Gamma $) plane for $c=0.3$ is shown in Fig. 4.

\section{Numerical simulations of the solitary pulses}

As it was said above, it is necessary to directly check whether stable SPs
indeed exist in the region where the stability is expected. To this end,
Eqs. (\ref{u}), (\ref{v}) with periodic boundary conditions (b.c.) were
integrated by means of an implicit Fourier spectral method
\cite{numeric}, the time step being, typically, $0.01$ and $0.02$.
(a description of the method is given in the
appendix). The initial conditions were taken as suggested by 
the perturbation theory, i.e., in the form (\ref{soliton}), but with
arbitrary values of the amplitude, in order to check whether strongly
perturbed pulses relax to stable ones, i.e., whether the stable pulses are 
\textit{attractors}.

Results are displayed in Fig. 1 by means of the symbols \textbf{x} and 
\textbf{o}, standing for unstable and stable solitons, respectively (it may
happen that, near the border with the unstable SPs, some pulses which appear
to be stable are subject to a very weak instability which does not manifest
itself within the integration time limits). As is seen, the pulses are
indeed stable everywhere inside the expected stability region. Moreover, all
the stable pulses were found to be strong attractors. For instance, in the
case $\alpha =0.1,\Gamma =0.15$, the initial pulses definitely relaxed to a
single stationary SP if their initial amplitude $A_{0}$ exceeded $1.7$. In
particular, starting with $A_{0}=3$ and $A_{0}=12$ at $t=0$, the pulse
attained the value of the amplitude, respectively, $A=6.91$ and $A=7.18$ at $%
t=400$. The analytical prediction for the amplitude of the steady-state
pulse (\ref{soliton}) (the one with the larger value of the amplitude) is,
at the same values of the parameters, $A_{\mathrm{anal}}\equiv 12\eta
^{2}\approx 6.45$, so that a discrepancy with the numerical results is less
than $10\%$. On the other hand, if the initial amplitude was too small,
e.g., $A_{0}=0.75$, the pulse decayed into zero, which is natural too, as
the zero solution has its own attraction basin. Note that for the second
(smaller) steady-state pulse, which is expected to play the role of the
separatrix between the attraction basins of the stable pulse and zero
solution, the perturbation theory predicts, in the same case, the amplitude $%
\widetilde{A}_{\mathrm{anal}}\approx 2.15$, so it seems quite natural that
the initial pulses with $A_{0}=3$ and $A_{0}=0.75$ relax, respectively, to
the stable pulse and to zero.

Figure 1 shows that the numerically found upper border of the stable-pulse
region is quite close to the border of the existence region for the
steady-state pulses, as predicted by the perturbation theory. Unlike this,
the numerically identified stability region extends far below the
analytically found border of the zero-solution instability. For instance, it
was found that, at $\alpha =0.15$ and $\Gamma =0.2$, when the zero solution
is unstable against perturbations with finite wavenumbers $k$, the fastest
growing perturbation corresponding to $k=k_{\max }\approx 1.3$, a\ fairly
stable pulse with the amplitude $A=11.75$ was found in the simulations, the
pulse's amplitude predicted by the perturbation theory being $A_{\mathrm{anal%
}}\approx 11.61$ in this case. Moreover, we ran simulations in which the
most dangerous perturbation with the above-mentioned wavenumber $k=1.3$ and
a rather large amplitude, $A_{\mathrm{pert}}=1$, was deliberately added to
the pulse in the initial configuration. Instead of growing and destroying
the pulse, the perturbation was \emph{suppressed} by the pulse, which
remained stable indefinitely long (Fig. 5).

A cause of this extended stability can be understood. A similar feature was
reported, for a generalized asymmetric (with respect to the reflection $%
x\rightarrow -x$) cubic-quintic GL equation with periodic b.c., which has
moving-pulse solutions, in Ref. \cite{Holland}. An explanation was that the
moving pulse, traveling across the integration domain, periodically passes
each point and suppresses the perturbation at a rate which exceeds the
perturbation growth rate (see also Ref. \cite{KSpulse2}, where stable pulses
were observed in the KS - KdV equation; in that work, an explanation was
that the moving pulse was able to escape growing perturbation wave packets).
It seems very plausible that a similar ``sweeping'' mechanism explains the
anomalous pulse stability in the present model. Indeed, when we repeated the
simulations for the same values of the parameters but in a spatial domain
four times as large (i.e., the corresponding sweeping rate is four times as
small), the pulse demonstrated the expected instability, even without any
specially added perturbation seed, see Fig. 6.

Similar to a result reported for a system of linearly coupled
Ginzburg-Landau (GL) equations in Ref. \cite{bound}, bound states (BSs) with
two or more peaks can be found in the present model, in addition to single
SPs. To this end, we took an initial configuration constructed as a set of
two or more identical exact solutions of the unperturbed system (\ref
{zero_order}) with a certain separation between them. The simulations have
shown that BSs featuring two or more peaks of equal amplitudes indeed
develop and propagate stably. These results for the two-peak and three-peak
BSs are illustrated by Figs. 7 and 8, respectively. We have also checked
that, even if the amplitudes of the initial pulses and separations between
them are changed, the same BS consisting of equidistant equal-amplitude
peaks finally develops, i.e., the BSs are fairly robust dynamical objects.
In this connection, it is relevant to mention that, in the above-mentioned
coupled GL equations, only two-pulse BSs are completely stable, while BSs of
three pulses are split by perturbations breaking their symmetry \cite{bound}.

\section{Conclusion}

In this work, we have introduced a model based on the Kuramoto - Sivashinsky
- Korteweg - de Vries equation,which is linearly coupled to an extra linear
dissipative equation. The model can be applied to the description of coupled
surface and interface waves\ on flowing multilayered liquid films. The
additional linear equation makes its possible to stabilize the zero
solution, which opens way to the existence of stable solitary pulses.
Treating the dissipation and gain in the model (but not the linear coupling
between the two wave modes) as small perturbations, and making use of the
balance equation for the net momentum, we have found that the condition of
the balance between the gain and dissipation may select two steady-state
solitons from their continuous family existing in the absence of the
dissipation and gain (the family was found in an exact analytical form).
When the zero solution is stable and, simultaneously, two SPs are picked up
by the balance equation for the momentum, the pulse with the larger value of
the amplitude is expected to be stable in the infinitely long system, while
the other pulse must be unstable, playing the role of a separatrix between
attraction domains of the stable pulse and zero solution. These predictions
have been completely confirmed by direct simulations. Moreover, if the
integration domain is not very large (and periodic boundary conditions are
imposed), some pulses are stable even when the zero background is unstable.
An explanation to the latter feature, based on the concept of periodic
suppression of the perturbations by the running pulse, is proposed.
Furthermore, stable bound states of two and three identical pulses have been
found numerically. An interesting issue, which is left for further work, is
a possibility of formation of stable periodic arrays of the pulses. Note
that periodic pulse arrays in the KS - KdV equation were studied in Refs. 
\cite{Takuji2,Takuji3}.

Lastly, it is worthy to note that, since we have found stable 
pulses in the model including the dispersion term, 
i.e., $u_{xxx}$ in Eq. (\ref{1}), a natural question is if 
the presence of this term is a necessary condition for the 
existence of stable pulses. Our preliminary numerical study 
demonstrates that there is a finite critical minimum value of the 
coefficient in front of the dispersive term
(if all the other parameters of the system are fixed) 
which is indeed necessary for the existence of stable pulses.
For instance, in a typical case considered in this work, with
$c = 0$, $\alpha =0.1$, $\gamma=0.05$, and $\Gamma=0.15$, this minimum
value was found to be roughly $0.5$.

A plausible qualitative explanation for this effect (which will be
considered in detail elsewhere) is that, although the dispersion 
does not directly affect the saturation mechanism stabilizing the 
pulses, it can inhibit mode coupling as an effective impedance, which
will lead to attaining the necessary saturation at a higher
amplitude of the pulse. Therefore, the amplitude 
of pulses becomes larger as the dispersion increases, while the pulse 
widths, directly determined by the dissipation, remain almost constant.
This trend to the formation of pulses with higher amplitudes in
the presence of strong dispersion is a feasible cause for the 
existence of stable pulses.

\section*{Acknowledgments}

This work was performed, in a part, within the framework of an exchange
program jointly supported by the Japan Society for Promotion of Science and
Israeli Ministry of Science and Technology. B.A.M. appreciates hospitality
of the Department of Aeronautics and Astronautics at the Kyoto University.
The authors appreciate a valuable discussion with G.I. Sivashinsky.

\section*{Appendix}

Here we give a detailed description of the numerical method employed for the
simulation of Eqs. (\ref{u}) and (\ref{v}). First, we introduce necessary
notation for the discrete Fourier transform (DFT). Suppose we work on the
interval $I=[0,L]$ with $L$-periodic functions. The interval is discretized
by means of a set of $N$ equidistant points, $x_{j}=jL/N$, where $%
j=0,1,\cdots ,N-1$, with the spacing between them $\Delta x=L/N$. At these
points, the numerical solutions for $u(x_j,t)$, $v(x_j,t)$ are denoted
by $u_j(t)$, $v_j(t)$ respectively. The corresponding spectral
variable is $\xi _{k}=2\pi k/L$ with $k\in \{-N/2,\cdots
,-1,0,1,\cdots ,N/2-1\}$ (actually, $N/2$ is an integer, see
below). Then DFT is given by  
\begin{equation}
\hat{u}_{k} = \mathcal{F}u_{j}=\sum_{j=0}^{N-1}u_{j}\exp 
\left( -i\xi _{k}x_{j}\right),\quad k=-\frac{N}{2},\cdots
,-1,0,1,\cdots ,\frac{N}{2}-1. 
\end{equation}

The inverse DFT is defined as 
\begin{equation}
u_{j}={\mathcal{F}}^{-1}\hat{u}_{k}=\frac{1}{N}\sum_{k=-N/2}^{N/2-1}\hat{u}%
_{k}\exp \left( i\xi _{k}x_{j}\right) ,\quad j=0,1,\cdots ,N-1.
\end{equation}
We employ the fast Fourier transform (FFT) to carry out the DFT and its
inverse in the numerical form, therefore $N$ must be taken equal to a powers
of 2.

The Fourier transform converts Eqs. (\ref{u}), (\ref{v}) into 
\begin{eqnarray}
\hat{u}_{t}+i\xi _{k}{\mathcal{F}}(\frac{1}{2}u^{2})-i\xi _{k}^{3}\hat{u}%
-i\xi _{k}\hat{v} &=&\alpha \xi _{k}^{2}\hat{u}-\gamma \xi _{k}^{4}\hat{u},
\label{num1} \\
\hat{v}_{t}+ic\xi _{k}\hat{v}-i\xi _{k}\hat{u} &=&-\Gamma \xi _{k}^{2}\hat{v}%
,  \label{num2}
\end{eqnarray}
where $k=-N/2,\cdots ,-1,0,1,\cdots ,N/2-1$.

For the time integration of Eqs. (\ref{num1}) and (\ref{num2}) , the
Crank-Nicolson scheme is used, which leads to a nonlinear system 
\begin{eqnarray}
\lefteqn{\frac{\hat{u}^{n+1}-\hat{u}^n}{\Delta t} + \frac 12 i \xi_k - i
\xi_k^3 \frac{\hat{u}^{n+1}+\hat{u}^n}{2} - i \xi_k \frac{\hat{v}^{n+1}+\hat{%
v}^n}{2}}  \nonumber \\
& + & \left({\mathcal{F}}(\frac{(u^n)^2}{2})
+{\mathcal{F}}(\frac{(u^{n+1})^2}{2})\right) =
(\alpha \xi_k^2-\gamma \xi_k^4) \frac{\hat{u}^{n+1}+\hat{u}^n}{2},
\label{eq:num3} \\
\lefteqn{\frac{\hat{v}^{n+1}-\hat{v}^n}{\Delta t} + i c\xi_k \frac{\hat{v}%
^{n+1}+\hat{v}^n}{2} - i \xi_k \frac{\hat{u}^{n+1}+\hat{u}^n}{2}}  \nonumber
\\
& = & -\Gamma \xi_k^2 \frac{\hat{v}^{n+1}+\hat{v}^n}{2}.  \label{eq:num4}
\end{eqnarray}

To solve this nonlinear system, the following iteration procedure is
employed: 
\begin{equation}
\hat{u}^{n+1,0} = \hat{u}^{n},\,\hat{v}^{n+1,0} = \hat{v}^{n},
\label{eq:num5}
\end{equation}

\begin{eqnarray}
\lefteqn{\frac{\hat{u}^{n+1,r+1}-\hat{u}^{n}}{\Delta t}-i\xi _{k}^{3}\frac{%
\hat{u}^{n+1,r+1}+\hat{u}^{n}}{2}-i\xi _{k}\frac{\hat{v}^{n+1,r+1}+\hat{v}%
^{n}}{2}}  \nonumber \\
&+&\frac{1}{2} i\xi _{k}\left({\mathcal{F}}(\frac{(u^{n})^{2}}{2})
+{\mathcal{F}}(\frac{%
(u^{n+1,r})^{2}}{2})\right) =(\alpha \xi _{k}^{2}-\gamma \xi _{k}^{4})\frac{%
\hat{u}^{n+1,r+1}+\hat{u}^{n}}{2},  \label{eq:num6} \\
\lefteqn{\frac{\hat{v}^{n+1,r+1}-\hat{v}^{n}}{\Delta t}+ic\xi _{k}\frac{\hat{%
v}^{n+1,r+1}+\hat{v}^{n}}{2}-i\xi _{k}\frac{\hat{u}^{n+1,r+1}+\hat{u}^{n}}{2}%
}  \nonumber \\
&=&-\Gamma \xi _{k}^{2}\frac{\hat{v}^{n+1,r+1}+\hat{v}^{n}}{2}.
\label{eq:num7}
\end{eqnarray}
Here $r=0,1,\cdots ,R-1$, where $R$ is the iteration number in each time
step. In practice, the number of the Fourier modes was taken to be $256$ or $%
512$, the typical time step was $0.01$ and $0.02$, and the iteration
was run twice in each time step. 

\newpage

\begin{center}
\textbf{Figure Captions}
\end{center}

Fig. 1. The stability region for solitary pulses in the parametric plane ($%
\alpha $,$\Gamma $) of the system (\ref{u}), (\ref{v}) for $\gamma =0.05$
and $c=0$. The zero solution is stable to the left of the dashed curve, and
Eq. (\ref{cubic}) produces two physical solutions to the right of the
continuous curve. The symbols \textbf{x} and \textbf{o} mark points at which
direct simulations show, respectively, that the solitary pulse is unstable
or stable.

Fig. 2. The expected stability region for the solitary pulses in the
parametric plane ($\alpha $,$\Gamma $) for $\gamma =0.05$ and $c=-1$. The
continuous and dashed curves have the same meaning as in Fig. 1.

Fig. 3. A typical result of an inelastic collision between two stable
solitons with different velocities, simulated within the framework of the
zero-order conservative system (\ref{zero_order}) with $c=0$: (a) $t=0$, (b) 
$t=0$, (c) $t=40$, (d) $t=40$, the initial velocities of the two solitons
are $s_{1}=4.236$ \ and $s_{2}=2.414$.

Fig. 4. The expected nearly vanishing stability region for the solitary
pulses in the parametric plane ($\alpha $,$\Gamma $) for $\gamma =0.05$ and $%
c=0.3$. The continuous and dashed curves have the same meaning as in Fig. 1.

Fig. 5. Suppression of the initially imposed large perturbation which is the
fastest growing instability mode in the infinite system by the traveling
pulse in the case $\alpha =0.15,\Gamma =0.2,\gamma =0.05$, and $c=0$, in the
spatial domain of the length $L=128$ with periodic boundary conditions. The
perturbation is taken as $u_{\mathrm{pert}}=v_{\mathrm{pert}}=a_{0}\cos
(k_{\max }x)$ with $a_{0}=1$ and $k_{\max }=1.3$. The panels (a) and (b)
show the initial configurations of the fields, and (c) and (d) are their
shapes produced by the direct simulations by the moment $t= 420$.

Fig. 6. The instability of the pulse at the same values of parameters and
for the same time interval as in Fig. 5, but in the spatial domain of the
length four times as large, $L=512$, without any specially imposed initial
perturbation.

Fig. 7. A stable bound state of two pulses found in the case $\alpha =0.1$, $%
\gamma =0.05$, $c=0$, $\Gamma =0.15$. The panels (a) and (b) show
established shapes of the $u$ and $v$ fields.

Fig. 8. A stable bound state of three pulses in the same case as in Fig. 7.

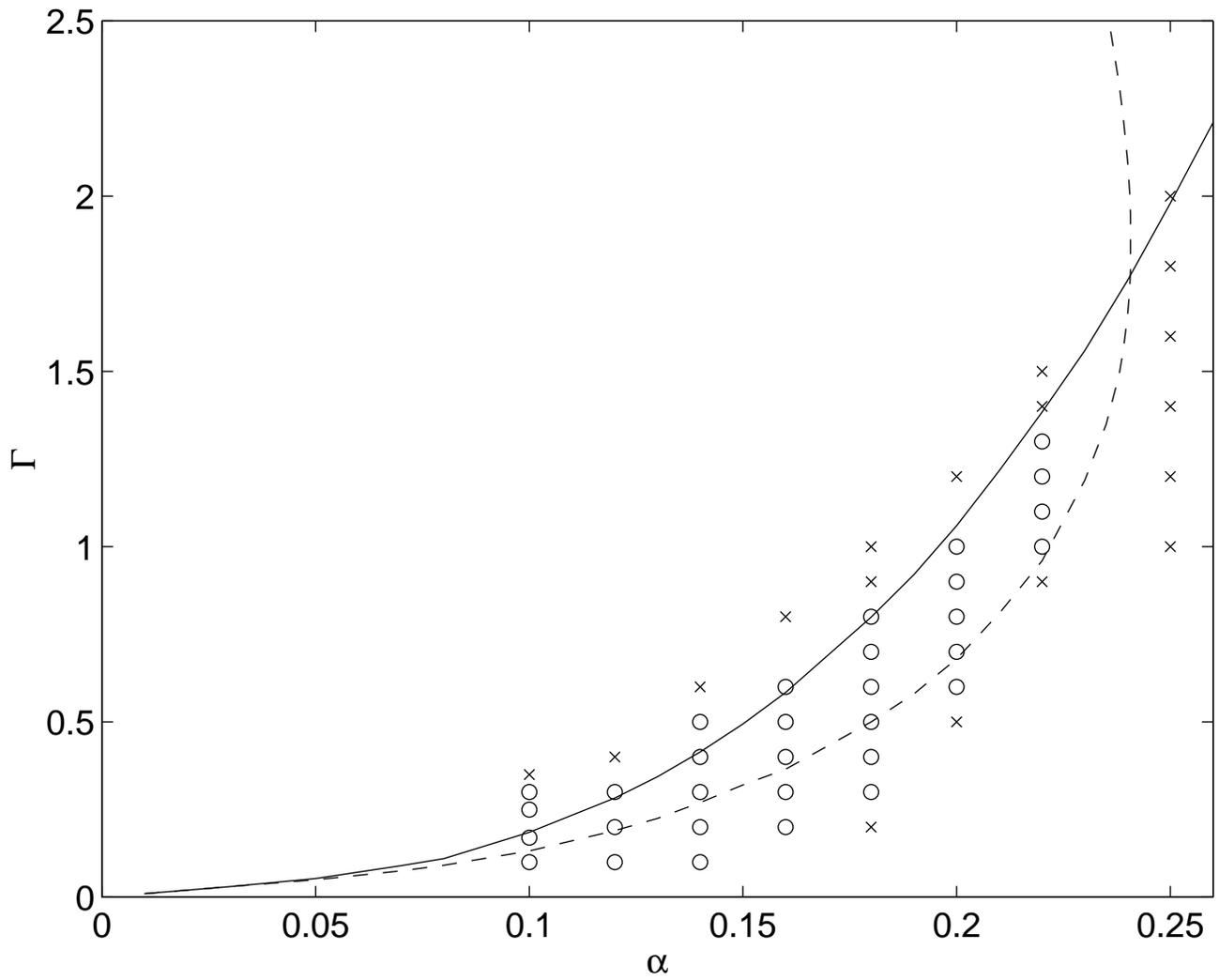
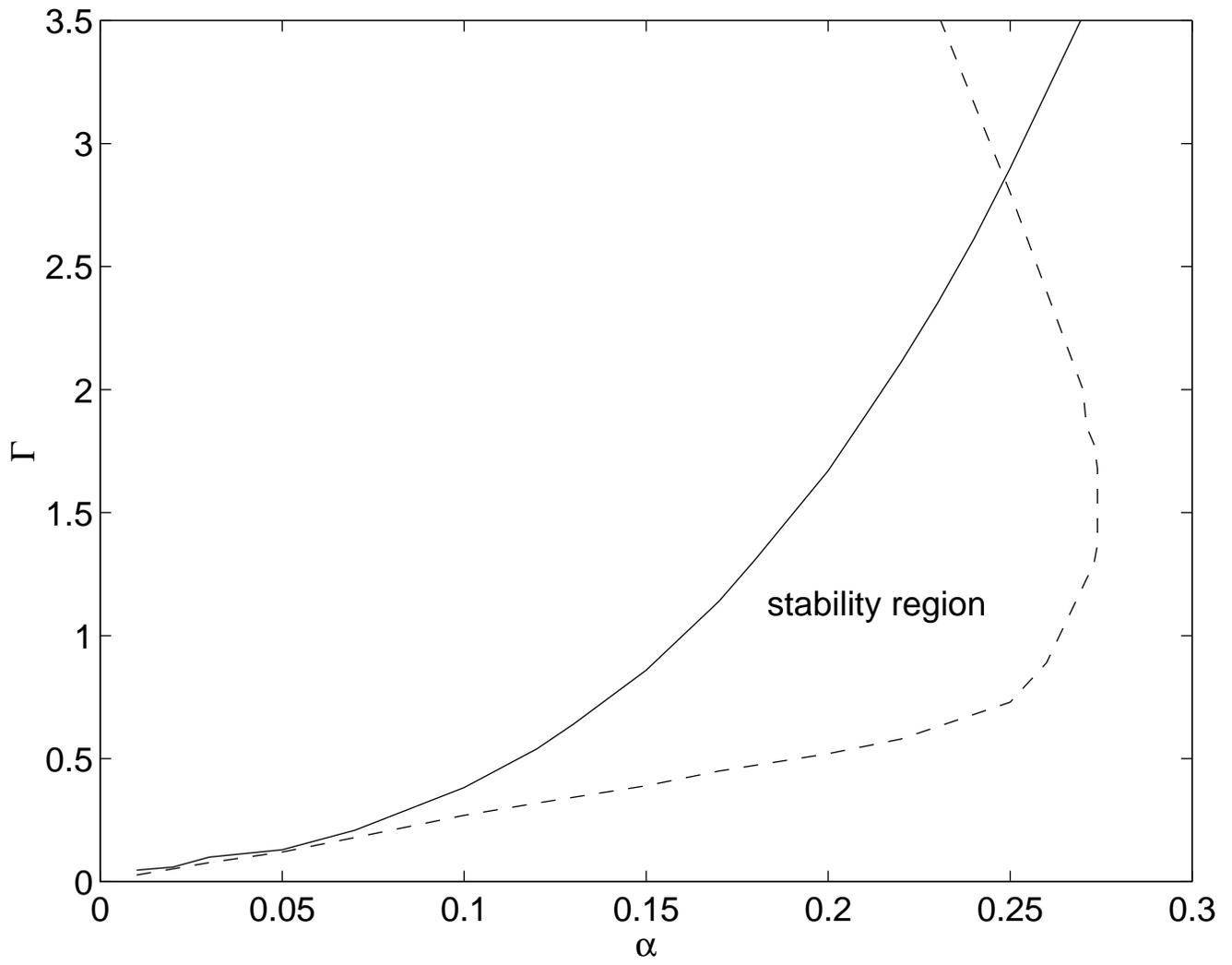
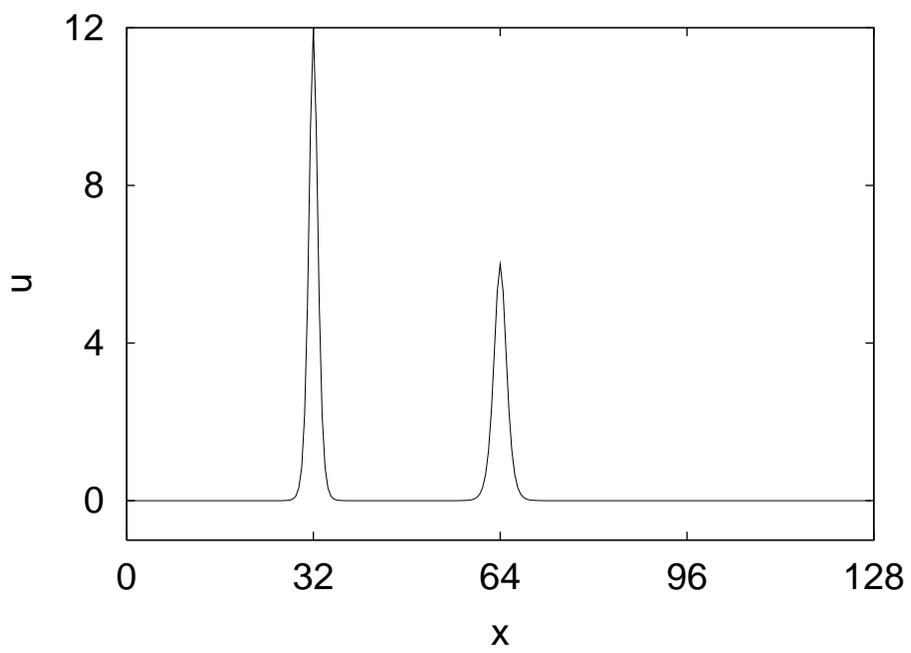
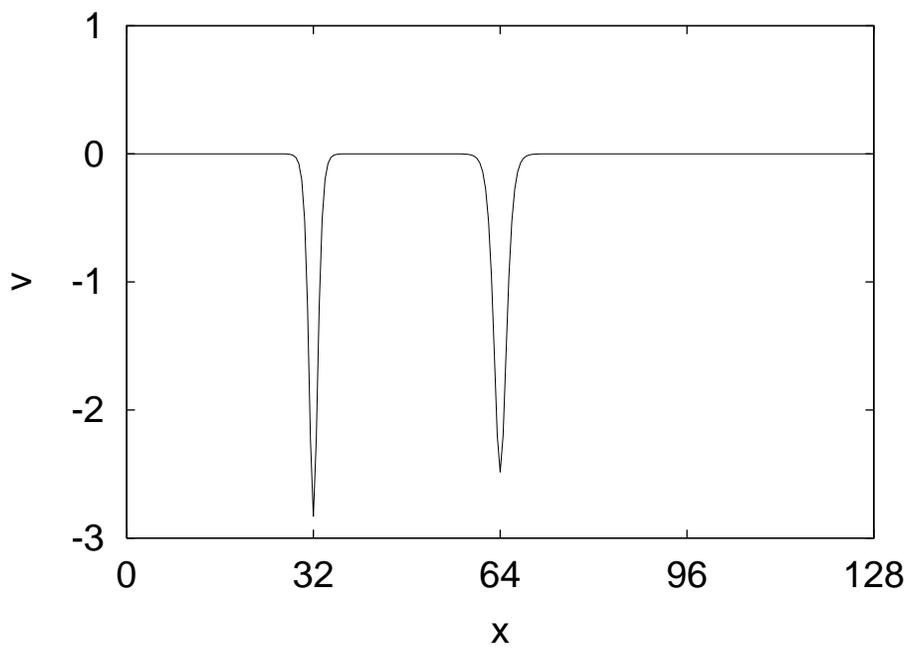
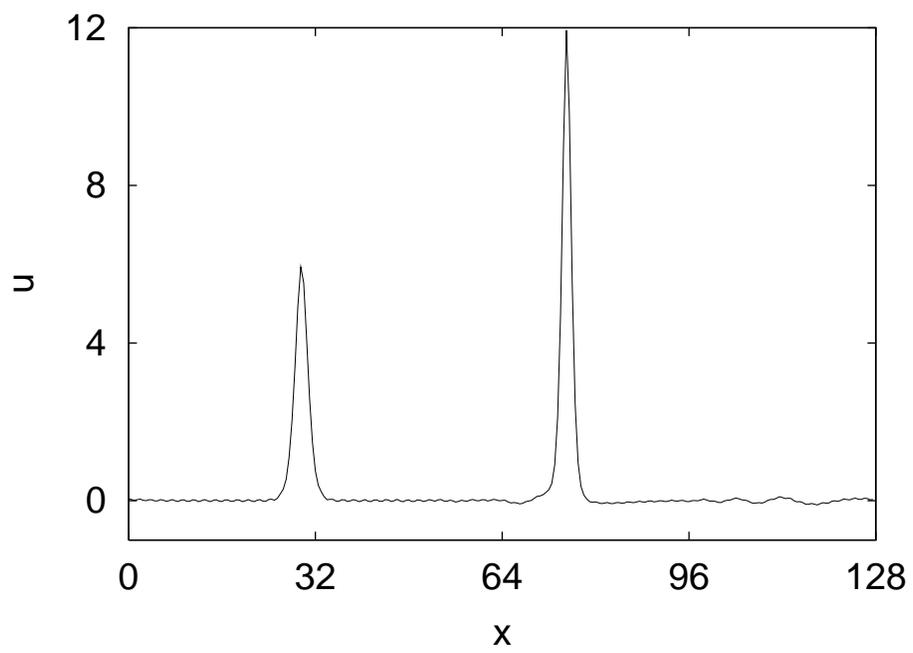
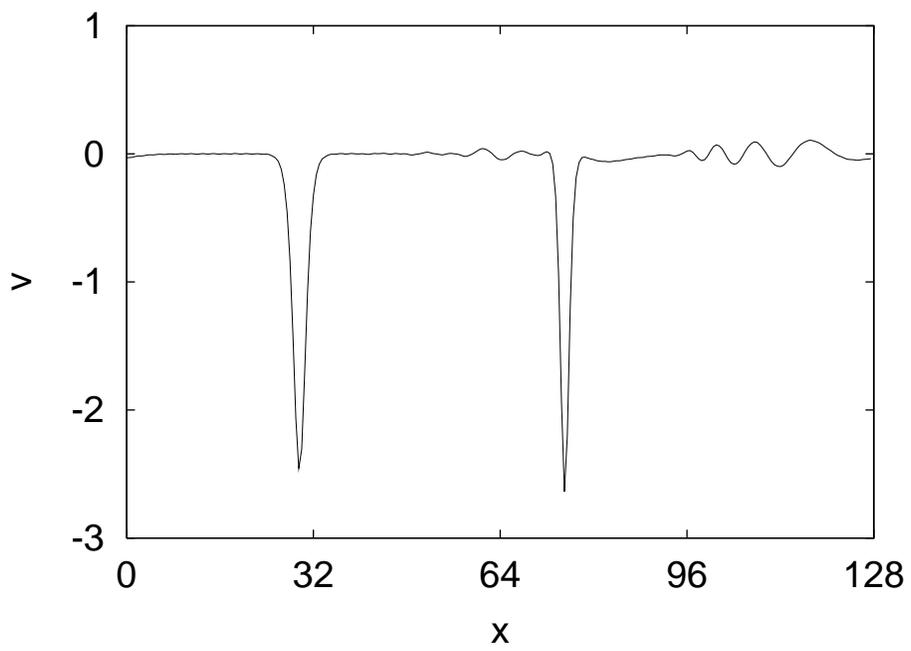
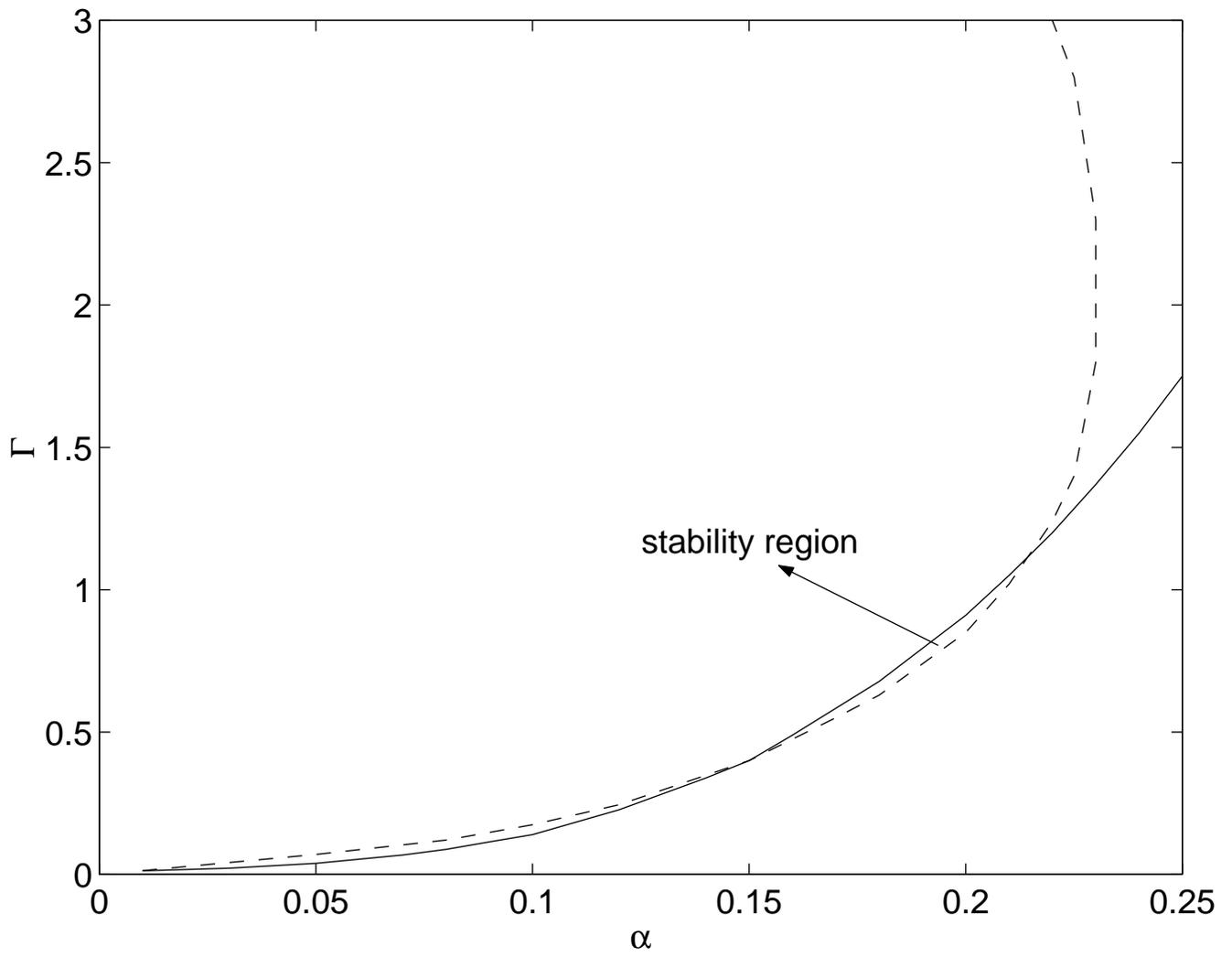
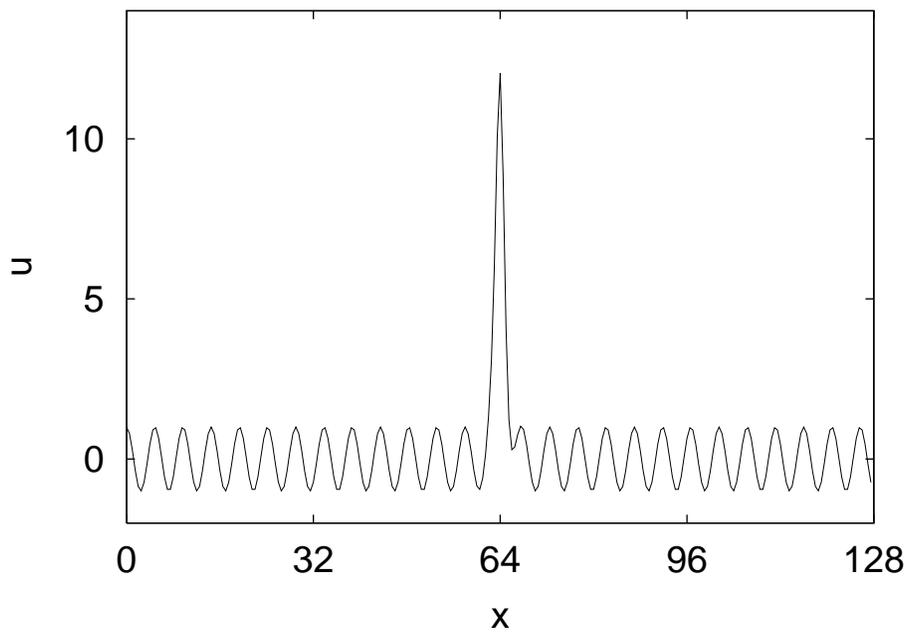
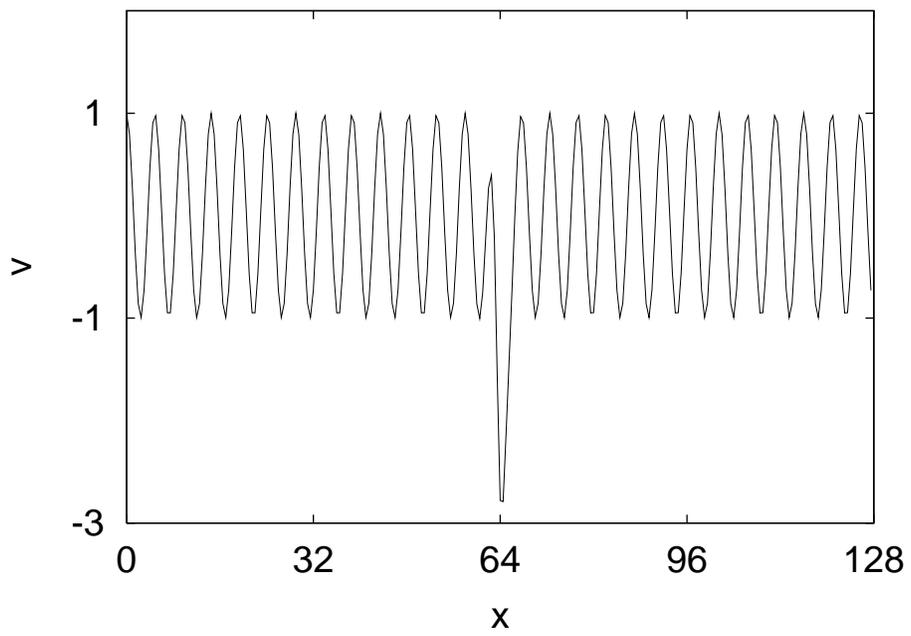
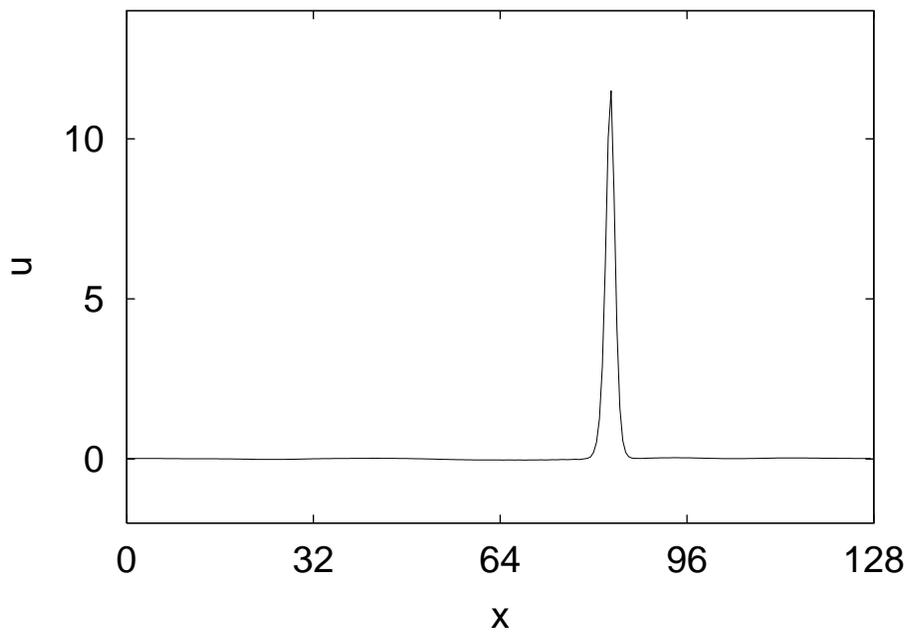
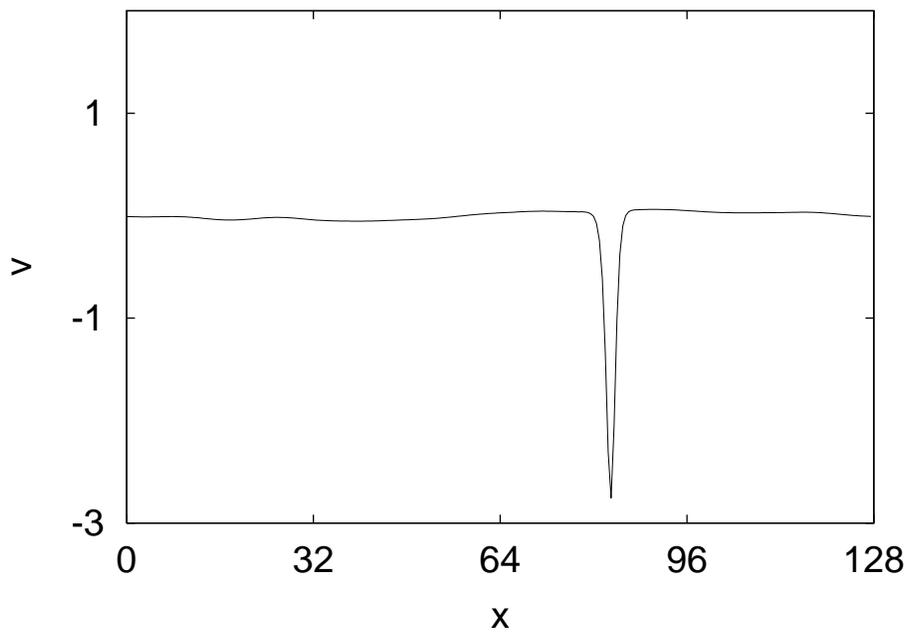
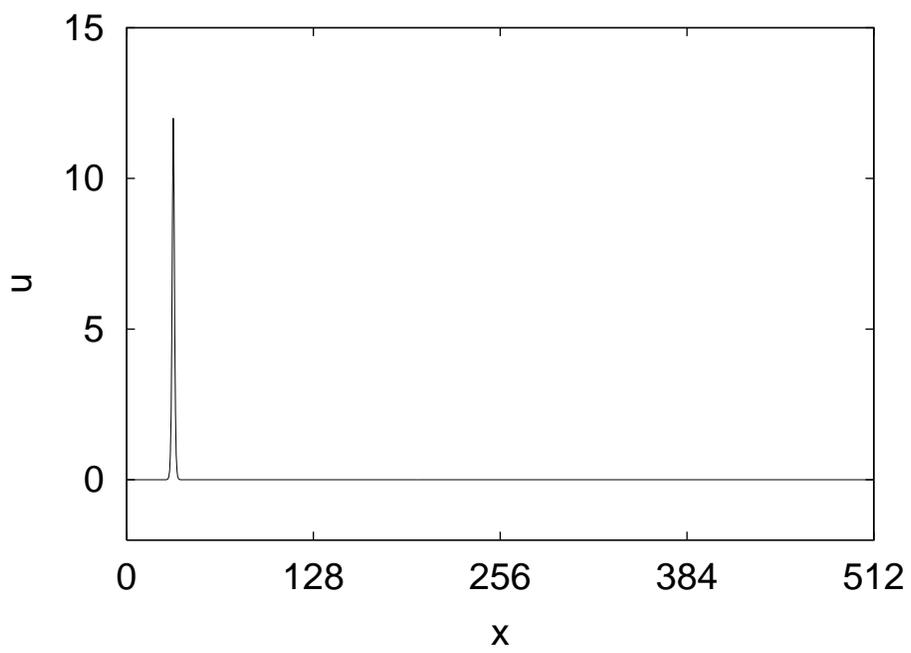
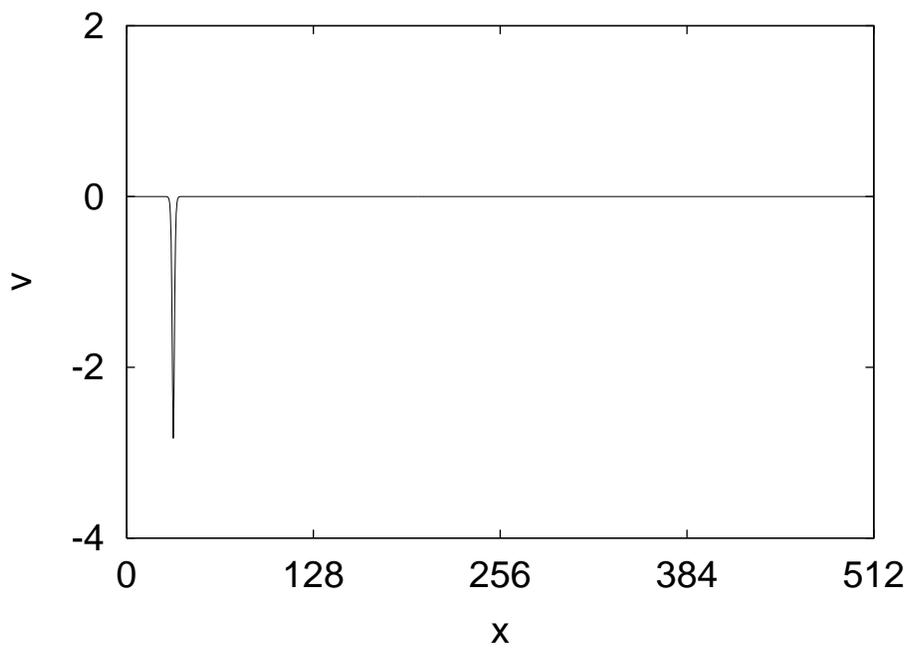
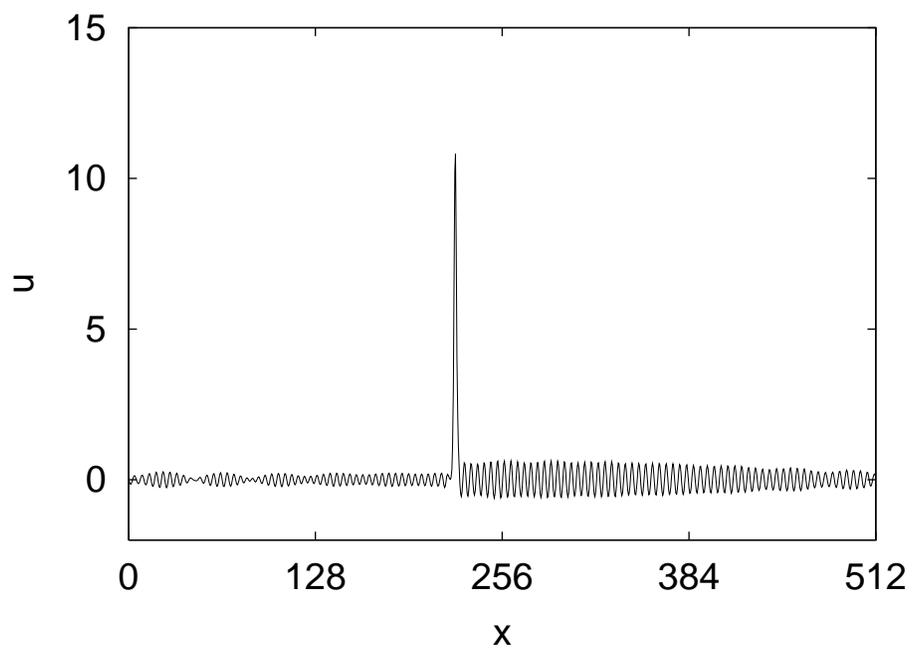
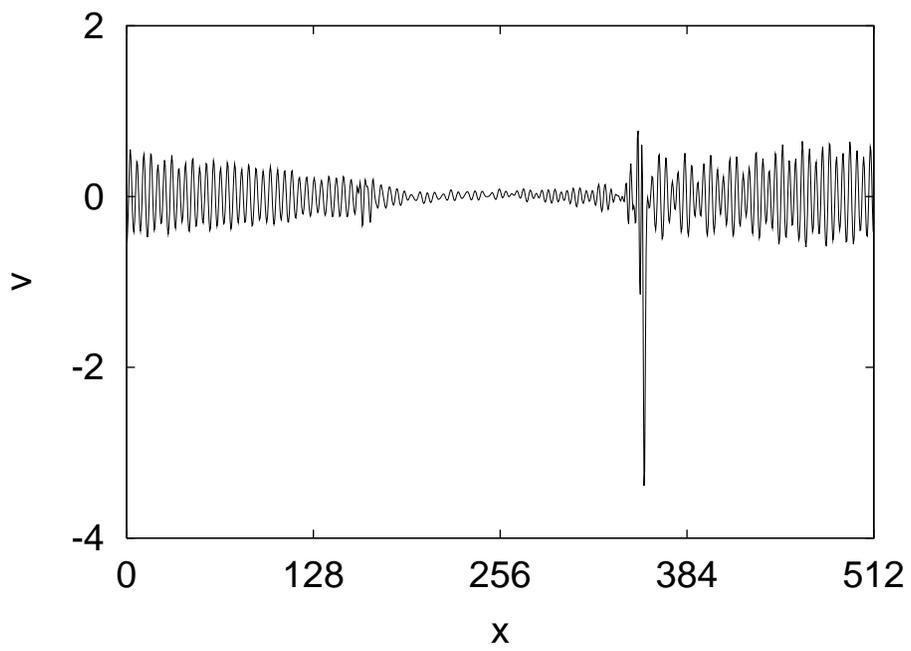
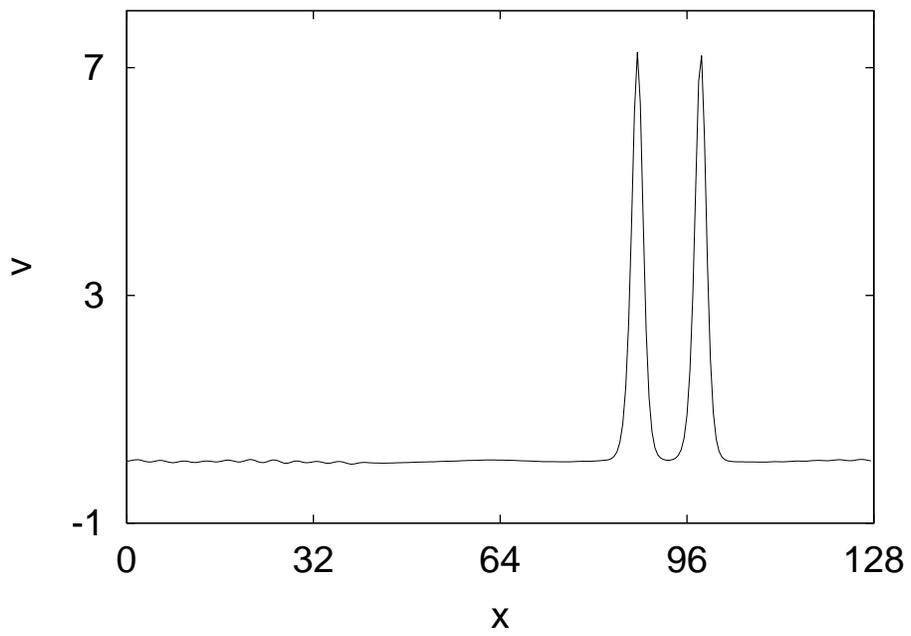
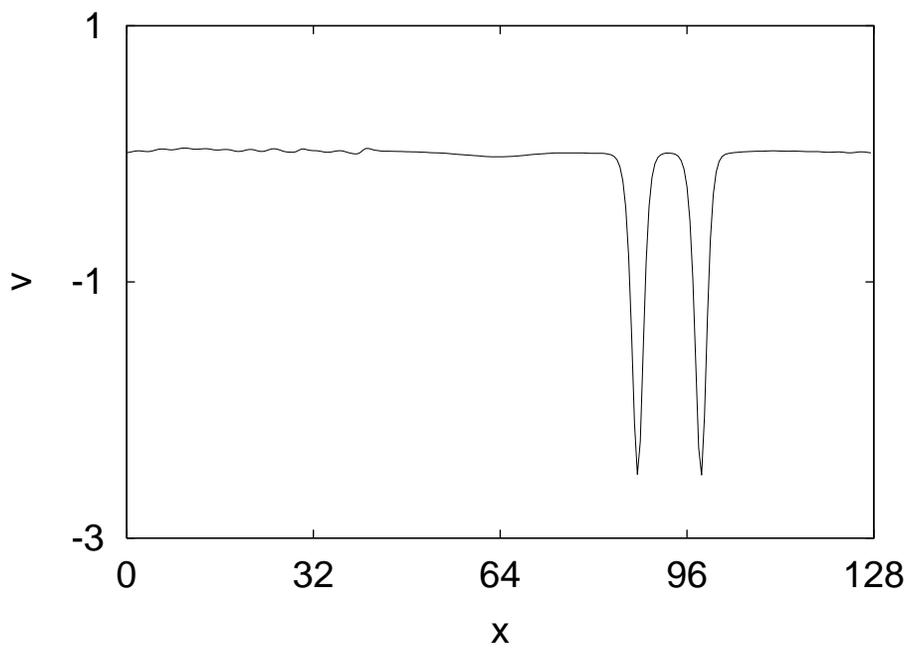
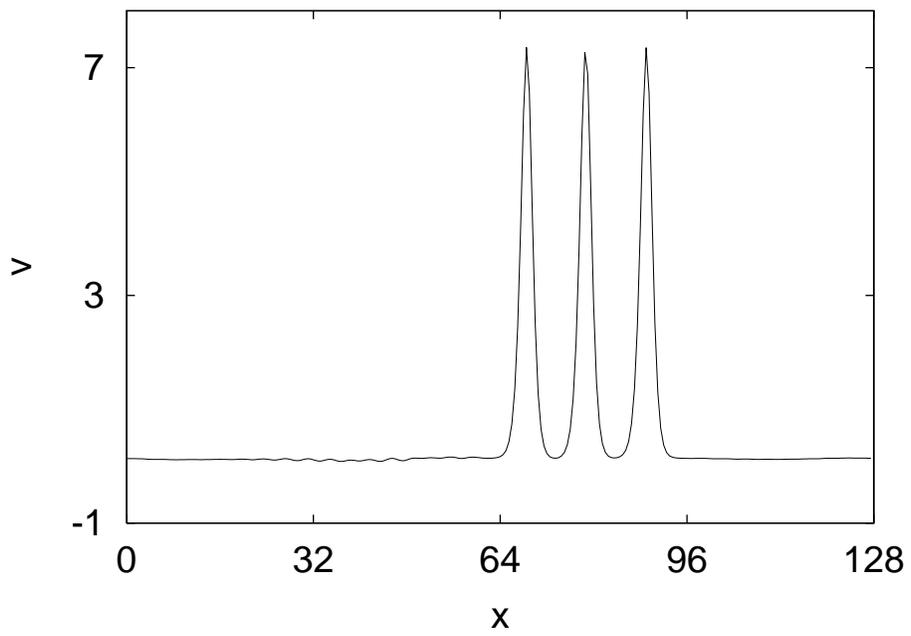
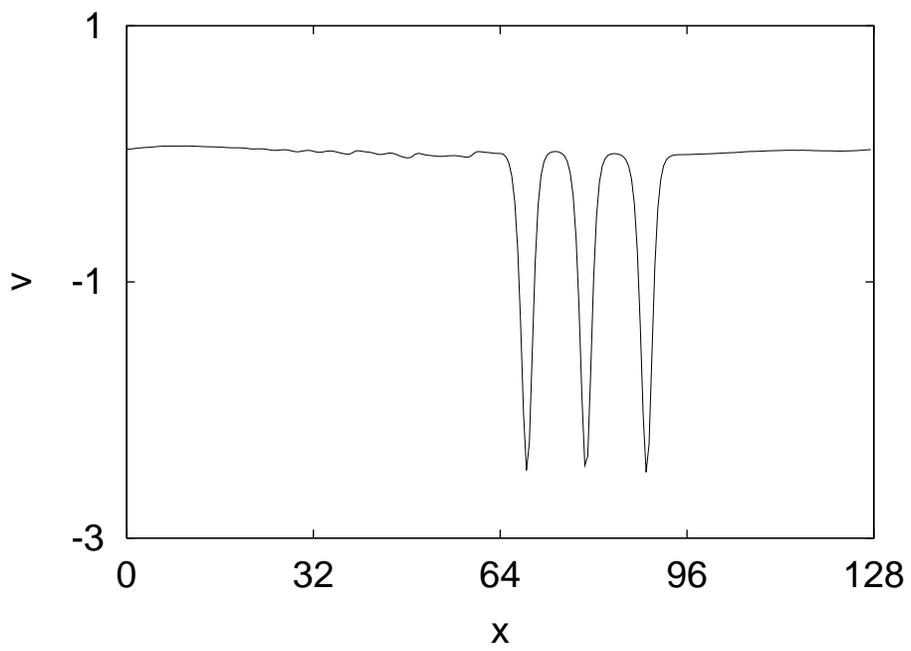
\end{document}